\documentclass{aa}

\usepackage{graphicx}
\usepackage{txfonts}
\def\lsim{\mathrel{\rlap{\lower4pt\hbox{\hskip1pt$\sim$}}
    \raise1pt\hbox{$<$}}}
\def\gsim{\mathrel{\rlap{\lower4pt\hbox{\hskip1pt$\sim$}}
    \raise1pt\hbox{$>$}}}
\newcommand{\beq}{\begin{equation}}
\newcommand{\eeq}{\end{equation}}

\def\ba{\begin{eqnarray}}
\def\ea{\end{eqnarray}}

\def\gs{\mathrel{\lower0.6ex\hbox{$\buildrel {\textstyle >}\over{\scriptstyle \sim}$}}}
\def\ls{\mathrel{\lower0.6ex\hbox{$\buildrel {\textstyle <}\over{\scriptstyle \sim}$}}}

\newcommand{\1}{\Omega_\mathrm{M}}
\newcommand{\2}{\Omega_\mathrm{r_c}}

\begin{document}

\title{Testing the DGP model with gravitational lensing statistics}

   \author{Zong-Hong Zhu\inst{1,2}
          \and
	Mauro Sereno\inst{3}
	}

   \offprints{Zong-Hong Zhu}

   \institute{Department of Astronomy, Beijing Normal
              University, Beijing 100875, China\\
              \email{zhuzh@bnu.edu.cn}
         \and
        Institut d'Astrophysique Spatiale (IAS),
        CNRS \& Univ. Paris-Sud,
        B\^atiment 121, F-91405 Orsay, France
        \and
        Institut f\"{u}r Theoretische Physik, Universit\"{a}t Z\"{u}rich,
	Winterthurerstrasse 190, CH-8057 Z\"{u}rich, Switzerland\\
        \email{sereno@physik.unizh.ch}
	}

\abstract
{}
{The self-accelerating braneworld model (DGP) seems to provide a simple 
 alternative to the the standard $\Lambda$CDM cosmology to explain the 
 current cosmic acceleration, which is strongly indicated by measurements 
 of Type Ia supernovae, as well as other concordant observations.} 
{In this work, we investigate observational constraints on this scenario
 from gravitational lensing statistics using the Cosmic Lens All-Sky Survey 
 (CLASS) lensing sample.}
{We show that a large parameter space of the DGP model is in good agreement
 with this radio source gravitational lensing sample.}
{In the flat case, $\Omega_\mathrm{K}=0$, the likelihood is maximized,
 ${\cal L}={\cal L_\mathrm{max}}$, for $\1 = 0.30_{-0.11}^{+0.19}$.
If we relax the prior on $\Omega_\mathrm{K}$, the likelihood peaks at
 $\{ \1,\2 \} \simeq \{0.29, 0.12\}$, just slightly in the region of open 
 models.
However the confidence contours are pretty elongated so that we can not 
 discard either close or flat or open models.}


\authorrunning{Zong-Hong Zhu and Mauro Sereno}
\titlerunning{Testing the DGP model with lensing statistics}
\maketitle

%

\section{Introduction}

The accelerating expansion of our universe was first discovered by the 
 measurements of distant Type Ia supernovae (SNe Ia; Riess et al. 1998; 
 Perlmutter et al. 1999), and was confirmed by the observations of the
 cosmic microwave background anisotropies (WMAP: Bennett et al. 2003)
 and the large scale structure in the distribution of galaxies (SDSS:
 Tegmark et al. 2004a,b).
By assuming General Relativity, a dark energy component has been invoked as 
 the most feasible mechanism for the acceleration. 
However, although fundamental for our understanding of the Universe, its 
 nature (as well as the nature of the dark matter) remains a completely open 
 question nowadays. 

Among several alternatives to dark energy, the models that make use of the 
 very ideas of branes and extra dimensions to obtain an accelerating universe
 are particularly interesting (Randall and Sundrum 1999a,b).
The general principle behind such models is that our 4-dimensional universe
 would be a brane embedded into a higher dimensional spacetime bulk on which
 gravity can propagate.
One famous brane world model is proposed by Dvali, Gabadadze and Porrati (2000),
 which is widely referred to as DGP model.
This scenario describes a self-accelerating 5-dimensional brane world model 
 with a noncompact, infinite-volume extra dimension in which the dynamics of 
 gravitational interaction is governed by a competition between a 4-dimensional
 Ricci scalar term, induced on the brane, and an ordinary 5-dimensional 
 Einstein-Hilbert action. 
For scales below a crossover radius $r_c$ (where the induced 4-dimensional 
 Ricci scalar dominates), the gravitational force experienced by two punctual 
 sources is the usual 4-dimensional $1/r^{2}$ force whereas for distance
 scales larger than $r_c$ the gravitational force follows the 5-dimensional 
 $1/r^{3}$ behavior.
The Friedmann equation is modified as follows 
\begin{equation}
\label{eq:ansatz}
H^2 = H_0^2 \left[
        \Omega_\mathrm{K}(1+z)^2+\left(\sqrt{\2}+
        \sqrt{\2+\1 (1+z)^3}\right)^2
                \right]
\end{equation}
where $H$ is the Hubble parameter as a function of redshift $z$ ($H_0$ is its
  value at the present), $\Omega_\mathrm{K}$, $\2$ and $\1$ represent
  the fractional contribution of curvature, the bulk-induced term and the
  matter (both baryonic and nonbaryonic), respectively.
$\2$ is defined as $\2 \equiv 1/4r_c^2H_0^2$.
From Eq.(1), the DGP model is a testable scenario with the same number 
 parameters as the standard $\Lambda$CDM model.

The advantages of the DGP model has triggered a wave of
 interests aiming to constrain its model parameters using various cosmological
 observations, such as
 the magnitude-redshift relation of supernovae of type Ia
	(Avelino and Martins 2002; 
         Deffayet et al. 2002; 
	Zhu and Alcaniz 2005; 
	Maartens and Majerotto 2006;
	Barger et al. 2007;
	Movahed et al. 2007),
 the cosmic microwave background shift parameter from WMAP and the baryon
 acoustic oscillation peak from SDSS
	(Guo et al. 2006;
	Lazkoz et al. 2006;
	Rydbeck et al. 2007;
	He et al. 2007), 
 the angular size - redshift data of compact radio sources
	(Alcaniz 2002),
 the age measurements of high-$z$ objects
	(Alcaniz, Jain and Dev 2002),
 the lookback time to galaxy clusters
	(Pires, Zhu and Alcaniz 2006),
 the optical gravitational lensing surveys 
	(Jain et al. 2002),
 the observed Hubble parameter $H(z)$ data 
	(Wan, Yi and Zhang) and
 the large scale structures 
	(Multam\"aki et al. 2003;
	Lue et al. 2004;
	Koyama and Maartens 2006;
	Song et al. 2007) 
 (For a recent review on the DGP phenomenology, see Lue 2006).

In this paper, we shall consider the observational constraints on the 
 parameters of the DGP model arising from the Cosmic Lens All-Sky Survey
 (CLASS) lensing sample. 
Our results are in agreement with other recent analyses, providing a
  complementary test to the DGP model.

Gravitational lensing has been becoming a useful tool for modern astrophysics.
It provides cosmological tests in several ways, such as 
 gravitational lensing statistics
	(Kochanek 1996; Zhu 1998; Cooray \& Huterer 1999; 
	Chiba and Yoshii 1999; Chae et al. 2002; Sereno 2005),
 weak lensing surveys
	(Benabed and Bernardeau 2001),
 Einstein rings in galaxy-quasar systems
	(Yamamoto \& Futamase 2001),
 clusters of galaxies acting as lenses on background high redshift galaxies
	(Sereno 2002; Sereno and Longo 2004; Sereno 2007),
 and gravitational lens time delay measurements (Schechter 2004).
Results from techniques based on gravitational lensing  are complementary to 
 other methods and can provide restrictive limits on the acceleration 
 mechanism.
The aim of the current paper is to check the validity of the DGP model with 
 radio-selected gravitational lensing statistics.
We adopt the Cosmic Lens All-Sky Survey (CLASS) statistical data which 
 consists of 8958 radio sources out of which 13 sources are multiply imaged 
 (Browne {\it et al.} 2003; Chae {\it et al.} 2002). 
We work only with those multiply imaged sources whose image-splittings are 
 known to be caused by single early type galaxies, which reduces the total 
 number of lenses to 10.
We show that a large parameter space of the DGP model is in good agreement
 with this radio source gravitational lensing sample.
The maximum likelihood happens at $\{ \1,\2 \} \simeq \{0.29, 0.12\}$, just
 slightly in the region of open models.

The paper is organized as follows.
In Section~2, the basics of gravitational lensing statistics is introduced.
Properties of the CLASS sample and its statistical analysis are illustrated
 in Section~3.
Finally, we present our conclusions and discussion in Section~4.


\section{Basics of gravitational lensing statistics}
\label{stat}

A realistic statistics of gravitational lenses can be performed based on 
 simple assumptions (Kochanek 1996; Chae 2003; Sereno 2005; 
 and references therein). 
The standard approach is based on the observed number count of galaxies and 
 on the simple singular isothermal sphere (SIS) model for lens galaxies.

The differential probability of a background source to be lensed by a 
 background galaxy with velocity dispersion between $\sigma$ and 
 $\sigma + d\sigma$ and in the redshift interval from $z_\mathrm{d}$ to 
 $z_\mathrm{d}+ d z_\mathrm{d}$ is
\beq
\label{stat1}
\frac{d^2 \tau}{d z_\mathrm{d} d \sigma} = \frac{d n_\mathrm{G}}{d \sigma}(z_\mathrm{d},\sigma) s_\mathrm{cr}(\sigma) \frac{cd t}{d z_\mathrm{d}} ,
\eeq
where $s_\mathrm{cr}$ is the cross section for lensing event and 
 $\frac{d n_\mathrm{G}}{d \sigma}$ is the differential number density of the 
 lens population. 
For a conserved comoving number density of lenses, 
 $n_\mathrm{G}(z) =n_0(1+z)^3$.

The lens distribution can be modeled by a modified Schechter function of the 
 form (Sheth et al. 2003)
\beq
\label{stat2}
\frac{d n_0}{d \sigma}=n_* \left( \frac{\sigma}{\sigma_*}\right)^\alpha \exp \left[ -\left( \frac{\sigma}{\sigma_*}\right)^\beta\right] \frac{\beta}{\Gamma (\alpha/\beta)} \frac{1}{\sigma},
\eeq
where $\alpha$ is the faint-end slope, $\beta$ the high-velocity cut-off and 
 $n_*$ and $\sigma_*$ are the characteristic number density and velocity 
 dispersion, respectively. 
Early-type or late-type populations contribute to the lensing statistics in 
 different ways and type-specific galaxy distributions  are required. 
As a conservative approach, we do not consider lensing by spiral galaxies. 
In fact the description of the late-type galaxy population is plagued by large 
 uncertainties and they contribute no more than 20-30\% of the total lensing 
 optical depth. 
A proper modeling of the distribution of the lensing galaxies is central in 
 lensing statistics. 
In our analysis we will use the results of Choi et al. (2007) who analyzed data 
 from the the SDSS Data Release 5 to derive the velocity dispersion 
 distribution function of early-type galaxies. 
They found $n_* = 8.0{\times} 10^{-3} h^3$~Mpc$^{-3}$, where $h$ is $H_0$ in 
 units of 100~km~s$^{-1}$~Mpc$^{-1}$, $\sigma_*=144 {\pm} 5$~km~s$^{-1}$, 
 $\alpha=2.49 \pm 0.10$, and $\beta = 2.29 \pm 0.07$.

Early-type galaxies can be well approximated as singular isothermal spheres. 
As shown in Maoz \& Rix (1993) and Kochanek (1996), radial mass distribution, 
 ellipticity and core radius of the lens galaxy are unimportant in altering 
 the cosmological limits. 
Assuming a flat model of universe, a typical axial ratio of 0.5 in a mixed 
 population of oblate and prolate spheroids would induce a shift of 
 $\sim 0.04$ in the estimation of $\1$ (Mitchell et al. 2005), well below 
 statistical uncertainties. 
Since departures from spherical symmetry induce a relatively small effect on
 lens statistics and the distribution of mass ellipticities is highly
 uncertain, spherically symmetric models supply a viable approximation. 
The cross section of a SIS is
\beq
\label{stat3}
s_\mathrm{cr}=16\pi^3 \left( \frac{\sigma}{c}\right)^4  \left( \frac{ D_\mathrm{d}  D_\mathrm{ds}}{D_\mathrm{s}} \right)^2,
\eeq
where $D_\mathrm{d}$, $D_\mathrm{ds}$ and $D_\mathrm{s}$ are the angular 
 diameter distances between the observer and the deflector, the deflector and 
 the source and the observer and the source, respectively. 
The two multiple images will form at an angular separation
\beq
\label{stat4}
\Delta \theta = 8 \pi \left( \frac{\sigma}{c}\right)^2  \frac{D_\mathrm{ds}}{D_\mathrm{s}} ,
\eeq
which relates the image separation to the velocity dispersion of the lens 
 galaxy. 
The total optical depth for multiple imaging of a compact source, $\tau$, the 
 probability that a SIS forms multiple images of a background source with 
 angular separation $\Delta \theta$, $d \tau /d\Delta \theta$, and the 
 probability of lensing by a deflector at 
 $z_\mathrm{d}$,  $d \tau/ d z_\mathrm{d}$, can be obtained by integrating 
 the differential probability in Eq.~(\ref{stat1}).

Lensing probabilities must be corrected for the magnification bias $B$, i.e. 
 the tendency of gravitationally lensed sources to be preferentially included 
 in flux-limited samples due to their increased apparent brightness 
 (Turner 1990; Fukugita \& Turner 1991; Fukugita et al. 1992; Kochanek 1993).
{\bf
The bias factor for a source at
redshift $z_{\rm s}$ with apparent magnitude $m$ is given by
\begin{eqnarray}
{\bf B}(m,z, M_0) & =&  \left(
\frac{dN_{\rm s}}{dm}\right)^{-1} \\
& {\times} & \int_{M_0}^{+\infty} \frac{dN_{\rm s}}{dm}(m+2.5\log M,z)P(M)dM , \nonumber
\end{eqnarray}
$M_0$ being the minimum magnification of a multiply imaged source,
with value $M_0 =2$; $P(M)dM =2 M_0^2M^{-3}dM$ is the probability that
a multiple image-lensing event causes a total flux increase by a
factor $M$ (Kochanek 1993). The function $dN_{\rm s}/d m$ is the
differential source number count in magnitude bins $dm$. 
}
Furthermore, since observations have finite resolution and dynamic range, 
 lens discovery rates are affected by the ability to resolve multiple source 
 images (Kochanek 1993). 
Lensing probabilities must then account for the resolution limit of the survey.
{\bf
For the SIS model,
selection effects can be characterized by the maximum magnitude
difference that can be detected for two images separated by $\Delta
\theta$, $\Delta m(\Delta \theta)$, which determines a minimum total
magnification $M_{\rm f} =M_0(f+1)/(f-1)$, where $2.5 \log f \equiv
\Delta m$ (Kochanek 1993).
}

Finally, the likelihood function can be written as (Kochanek 1993; 
 Chae et al. 2002)
\beq
{\cal L}= \prod_{i=1}^{N_{\rm U}}(1-p_i)\prod_{j=1}^{N_{\rm
L}}p_{l,j},
\eeq
where $N_{\rm L}$ is the number of multiple-imaged sources and $N_{\rm U}$ is 
 the number of unlensed sources. 
$p_l$ is the suitable probability accounting for the whole of the data 
 available for each lens system, i.e. the lens redshift and/or the
 image separation (Chae et al. 2002; Mitchell et al. 2005).
Probabilities are corrected for bias and selection effects.

Since $\tau \ll 1$ the likelihood can be approximated as 
 (Mitchell et al. 2004) 
\beq
{\cal L} \simeq \exp \left[ -\int N_z(z_\mathrm{s})p (z_\mathrm{s})d
z_\mathrm{s}
\right]
\prod_{j=1}^{N_{\rm L}}p_{l,j},
\eeq
where $N_z(z_\mathrm{s})$ is the redshift distribution of the sources. 
We use a uniform distribution for the priors on the cosmological parameters, 
 so that, apart from an overall normalization factor, the likelihood can be 
 identified with the posterior probability.


\section{Data analysis}

In this section, we discuss the radio-survey used for our lensing statistics 
  and present the constraints on the parameters of the DGP model.

\subsection{Data set}

The most reliable data set suitable for statistical analysis is provided by 
 a sample of 8958 flat-spectrum radio sources with 13 lenses by the Cosmic 
 Lens All-Sky Survey (CLASS; Browne et al. 2003; Myers et al. 2003). 
Data of interest are listed in table~1 of (Chae 2005). 
We limit our analysis to the early-type lens galaxies. 
Ten systems in the CLASS sample (0445+123, 0631+519, 0712+472, 1152+199, 
 1359+154, 1422+231, 1608+656, 1933+503, 2114+022 and 2319+051) can be 
 assumed to be early-type lenses (Chae 2005). 
We do not consider the information on the image separation in
 1359+154, 1608+656 and 2114+022 
 whose splittings are strongly affected by galaxy companions very close to
 the main lens.

The final CLASS statistical sample has been selected such that, for doubly 
 imaged systems, the flux ratio is $\leq 10$ and it is independent of the 
 angular separation. 
According to the selection criteria, the compact radio-core images have 
 separations greater than $ \Delta \theta_{\rm min}=0.3$~arcseconds. 
The probabilities that enter the likelihood must be then considered as the 
 probabilities of producing image systems with separations $\geq \Delta 
 \theta_{\rm min}$. 
Taking into account the CLASS observational selection function, 
 Chae (2007) found a magnification bias of $B \simeq 3.36$ for the SIS.

Redshift measurements are only available for a restricted CLASS subsample. 
Following Sereno (2005), we model the redshift distribution  
 $N_z(z_\mathrm{s})$ of the sources with a kernel empirical estimator. 
For the unmeasured lensed source redshifts, we set $z_\mathrm{s}$ to the 
 mean redshift of the sources lensed by early-type galaxies with measured 
 redshift, $\langle z_\mathrm{s} \rangle_{\rm lensed}=2.2$.

\subsection{Statistical analysis}

\begin{figure}
   \centering
   \includegraphics[width=8.9cm]{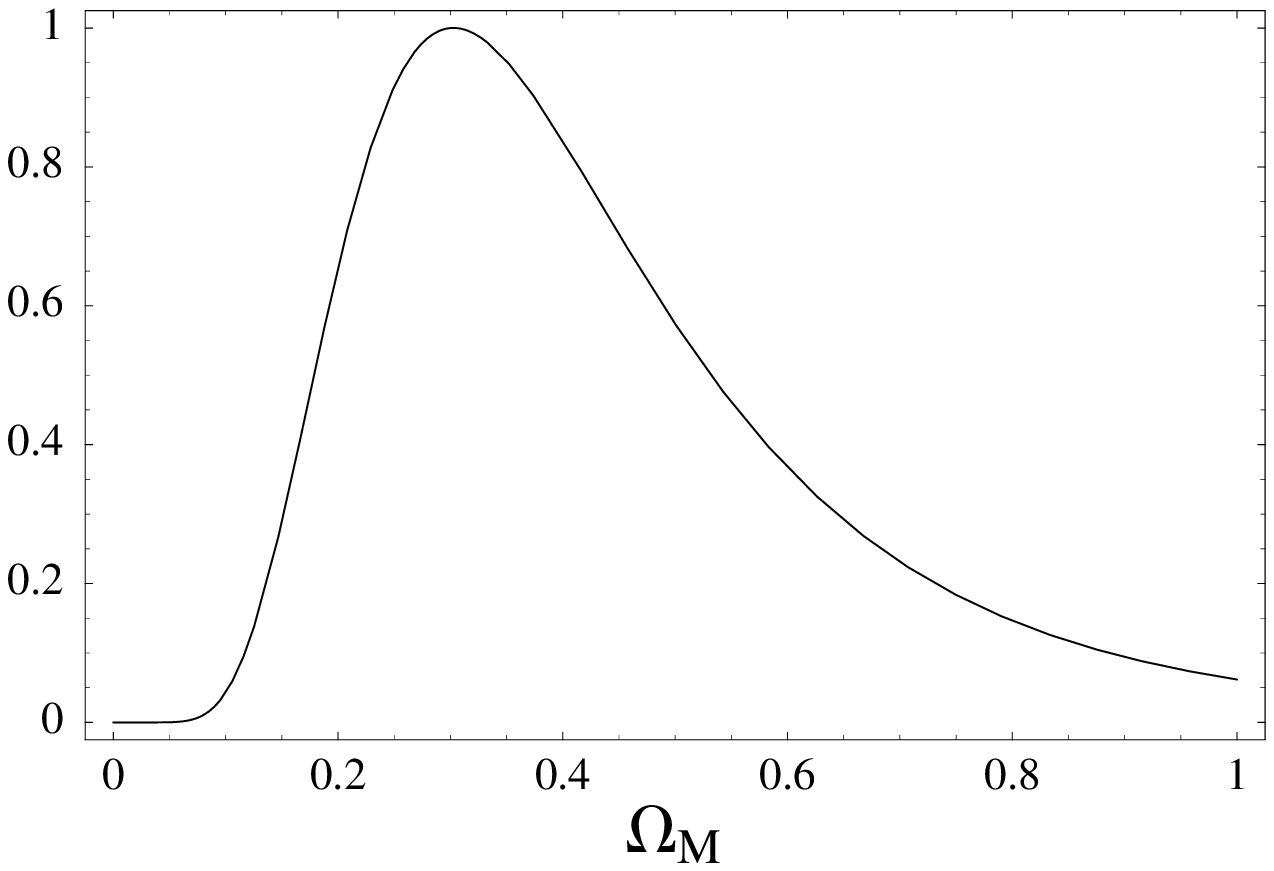}
\caption{Normalized likelihood, ${\cal L}/{\cal L}_\mathrm{max}$, as a
        function of $\1$ for a flat geometry, $\Omega_\mathrm{K}=0$.}
\label{Like_1Par}
\end{figure}

Let us now perform a statistical analysis of the data sample. 
As a first step, we fix the nuisance galactic parameters to their central 
 values. 
We will consider the related uncertainty later. 
In the flat case, $\Omega_\mathrm{K}=0$, the likelihood is maximized, 
 ${\cal L}={\cal L_\mathrm{max}}$, for $\1 = 0.30_{-0.11}^{+0.19}$, 
 see Fig.~\ref{Like_1Par}. 
Uncertainties denote the statistical 68.3\% confidence limit for one 
 parameter, determined by ${\cal L}/{\cal L_\mathrm{max}}=\exp (-1/2)$.

\begin{figure}
   \centering
   \includegraphics[width=8.9cm]{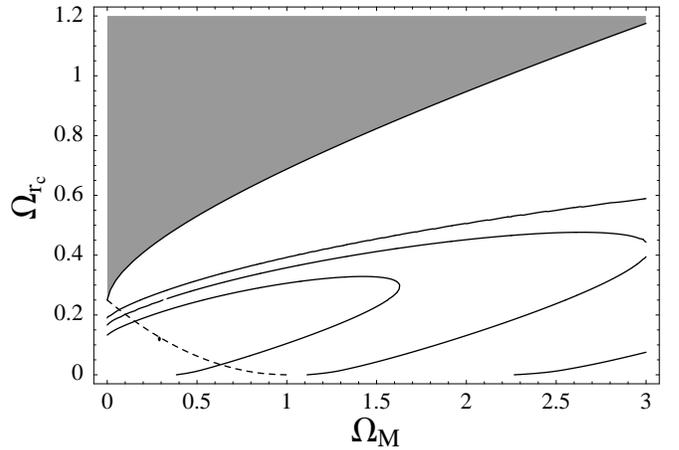}
\caption{Normalized likelihood, ${\cal L}/{\cal L}_\mathrm{max}$, in
        the $\1$-$\2$ plane. The dot shows the best fit model and the
        contours denote the $68.3\%$, $95.4\%$ and $99.7$ confidence limits
        for two parameters. The dashed line represents the locus of flat
        models of universe ($\Omega_\mathrm{K}=0$); bouncing models in the
        upper-left shaded region do not have big bang.}
\label{Like_OM_Orc}
\end{figure}

Even if we relax the prior on $\Omega_\mathrm{K}$, the likelihood peaks for 
 nearly flat models. 
In fact, the likelihood is maximum for $\{ \1,\2 \} \simeq \{0.29, 0.12\}$, 
just slightly in the region of open models, see Fig.~\ref{Like_OM_Orc}. 
The three contours in the figure correspond to the $68.3\%$, $95.4\%$ 
 and $99.7\%$ confidence limits for two parameters, 
 namely ${\cal L}/{\cal L_\mathrm{max}}=\exp (-2.30/2)$ , $\exp (-6.17/2)$ 
 and $\exp (-11.8/2)$, respectively. 
However, contours are pretty elongated so that we can not discard either 
 close or flat or open models.

Uncertainties in the redshift distribution of the sources can induce 
 additional errors in the estimates of the cosmological parameters. 
A source of error is the finite sample size of the sample of measured 
 source  redshifts (only 27 source redshifts are known), which induces an 
 error in the estimated redshift distribution. 
From a bootstrap resampling procedure, it can be created a set of simulated 
 distributions which is then used to create a new kernel estimator for the 
 redshift distribution. 
It can be shown that the finite size induces a dispersion of $\sim 0.08$ 
 on $\1$ (Sereno 2005). 
On the other hand, the cosmological constraints are nearly insensitive of 
 the functional form used when modeling the redshift distribution. 
Conclusions are really unaffected if a Gaussian distribution is used 
 instead of the kernel estimator. 
Finally, results change in a very negligible way if we use different values 
 of $z_\mathrm{s}$ for the lensed sources with unknown redshift.

The main uncertainty in the estimation of cosmological parameters comes 
 from errors in the assumed parameters of the velocity dispersion 
 distribution function which describes the lens population. 
In order to estimate such source of error, we simulated a sample of 100 
 sets of galactic parameters by extraction from normal distributions 
 centered on the best estimates of each parameter and with standard 
 deviation given by the associated uncertainty. 
The likelihood analysis was then repeated for each set of galactic parameters. 
Assuming flat cosmological models, the resulting distribution of the maximum 
 likelihood estimates has a scatter of $\sim 0.09$, which gives a similar 
 uncertainty in the determination of $\1$.

Finally a theoretically important systematic uncertainty is due to the the 
 effect of small-scale inhomogeneities on large-scale observations. 
Matter distribution is locally inhomogeneous and affects light propagation 
 and the related cosmological distances 
 (Sereno et al. 2001; Sereno, Piedipalumbo and Sazhin 2002; 
  and references therein).
However, being the universe globally homogeneous, the effect on the total 
 lensing statistics is small (Covone et al. 2005).

\section{Conclusion and discussion}

Since the discovery of the accelerating expansion of the universe, 
 in addition to the standard $\Lambda$CDM cosmological model, a huge
 number of scenarios have been proposed to be the acceleration mechanism
 (for a recent review, see: Sahni and Starobinsky 2000; Padmanabhan 2003; 
 Lima 2004; Copeland, Sami and Tsujikawa 2006; Alcaniz 2006).
Examples include 
  the so-called ``X-matter"
        (Turner and White 1997;
        Zhu, Fujimoto and Tatsumi 2001;
        Alcaniz, Lima and Cunha 2003;
        Dai, Liang and Xu 2004;
	Rupetti et al. 2007;
	Wang, Dai and Zhu 2007),
  a decaying vacuum energy density or a time varying $\Lambda$-term
        (Ozer and Taha 1987; Vishwakarma 2001),
  an evolving scalar field, dubbed quintessence
        (Ratra and Peebles 1988;
        Caldwell et al. 1998;
        Wang and Lovelace 2001;
        Gong 2002;
        Chen and Ratra 2004;
	Choudhury and Padmanabhan 2005;
	Ichikawa et al. 2006),
  the phantom energy, in which the sum of the pressure and energy
    density is negative
        (Caldwell 2002;
        Dabrowski et al. 2003;
        Wang, Gong and Su 2004;
	Wu and Yu 2005, 2006;
	Chang et al. 2007),
  the Chaplygin gas
        (Kamenshchik et al. 2001;
        Bento et al. 2002;
        Alam et al. 2003;
        Alcaniz, Jain and Dev 2003;
        Dev, Alcaniz and Jain 2003;
        Silva and Bertolami 2003;
        Makler et al. 2003;
	Zhu 2004;
	Zhang and Zhu 2006),
  the quintom model
	(Feng, Wang and Zhang 2005;
	Guo et al. 2005;
	Zhao et al. 2005; Xia et al. 2006;
	Wei and Zhang 2007),
  the holographic dark energy
	(Li 2004; Zhang and Wu 2005; Chang, Wu and Zhang 2006),
  the Cardassion model
        (Freese and Lewis 2002;
        Zhu and Fujimoto 2002, 2003;
        Sen and Sen 2003;
        Wang et al. 2003;
        Gong and Duan 2004a,b;
	Wang 2005;
	Bento et al. 2006;
	Reboul and Cordoni 2006;
	Yi and Zhang 2007)
  and the Casimir force (Szydlowski and Godlowski 2007; Godlowski et al. 2007).
All these acceleration mechanisms should be tested with various astronomical
 observations.

In this paper, we have focused our attention on the DGP model. 
We have analyzed this scenario by using the Cosmic Lens All-Sky Survey 
 sample (Browne et al. 2003; Myers et al. 2003) to obtain the 68.3\%, 95.4\%
 and 99.7\% confidence regions on its parameters.
It is shown that a large parameter space of the DGP model is consistent with
 this radio source gravitational lensing sample.
In the flat case, $\Omega_\mathrm{K}=0$, the likelihood is maximized,
 ${\cal L}={\cal L_\mathrm{max}}$, for $\1 = 0.30_{-0.11}^{+0.19}$.
If we relax the prior on $\Omega_\mathrm{K}$, the likelihood peaks at
 $\{ \1,\2 \} \simeq \{0.29, 0.12\}$, just slightly in the region of open 
 models.
The obtained confidence regions of Figure~2 are also in good agreement with 
 the results from analyzing data of type Ia supernovae (Zhu and Alcaniz 2005),
 which implies that gravitational lensing statistics provides an 
 independent and complementary constraint on the DGP model.
However, similar to the case of type Ia supernovae, 
 the confidence contours are pretty elongated so that we can not
 discard either close or flat or open models by only using the CLASS sample.
{\bf
Using the \emph{gold} sample of type Ia supernovae (SNeIa), the first year 
data from the Supernova Legacy Survey (SNLS) and the baryon acoustic 
oscillation (BAO) peak found in the Sloan Digital Sky Survey (SDSS),
Guo et al (2006) obtained, at 99.73\% confidence level, 
 $\Omega_m=0.270^{+0.018}_{-0.017}$ and $\Omega_{r_c}=0.216^{+0.012}_{-0.013}$
 (hence a spatially closed universe with $\Omega_k=-0.350^{+0.080}_{-0.083}$),
 which seems to be in contradiction with the most recent WMAP results 
 indicating a flat universe.
Based on this result, the authors also estimated the transition redshift 
 (at which the universe switches from deceleration
 to acceleration) to be $0.70 < z_{q=0} < 1.01$, at $2\sigma$ confidence level.
Therefore, the method of combining observational data provides much more 
 stringent constraint on the DGP model than any single data.
}
It is naturally hopeful that, with either future larger gravitational lensing
 samples or a joint investigation with other astronomical observations, one
 could obtain a more stringent constraint on the DGP model parameters.

\begin{acknowledgements}
This work was supported by
  the National Natural Science Foundation of China, under Grant No. 10533010,
  973 Program No. 2007CB815401, Program for New Century Excellent Talents in
  University (NCET) of China
  and the Project-sponsored by SRF for ROCS, SEM of China.
Z.-H. Z. also acknowledges support from CNRS,
  and is grateful to all members of cosmology group at IAS for their
  hospitality and help during his stay.
M.S. thanks the Department of Astronomy, Beijing Normal University, for the 
 warm hospitality and the financial support during its visit in Beijing. 
M.S. is supported by the Swiss National Science Foundation and by the Tomalla 
 Foundation.
\end{acknowledgements}

\end{document}